\documentclass[aps,twocolumn,prl,showpacs,10pt]{revtex4-1}

\usepackage{ifpdf}
\usepackage{enumerate}

\ifpdf 
	\usepackage[pdftex]{graphicx} 
\else
	\usepackage[dvips]{graphicx} 
\fi

\usepackage{amsmath}
\usepackage{amsfonts}
\usepackage{amssymb}
\usepackage[sort&compress]{natbib}
\usepackage{verbatim} 
\usepackage{bm}

\begin{document}
\title{Recolliding periodic orbits}

\author{A. Kamor$^{1,2}$, F. Mauger$^{2}$, C. Chandre$^2$, and T. Uzer$^1$}

\affiliation{$^1$ School of Physics, Georgia Institute of Technology, Atlanta, GA 30332-0430, USA\\
$^2$ Centre de Physique Th\'eorique - UMR 7332, CNRS -- Aix-Marseille University, Campus de Luminy, case 907, 13009 Marseille, France}

\begin{abstract}
We show that a family of key periodic orbits drive the recollision process in a strong circulary polarized laser field.  These orbits, coined recolliding periodic orbits, exist for a wide range of parameters and their relative influence changes as the laser and atomic parameters are varied.  We find the necessary conditions for recollision-driven nonsequential double ionization to occur.  The outlined mechanism is universal in that it applies equally well beyond atoms:  The internal structure of the target species plays a minor role in the recollision process.
\end{abstract}

\pacs{32.80.Rm, 05.45.Ac, 33.80.Rv}

\maketitle
Can an ionized electron be driven back to the core by an ultrastrong laser pulse? This is a high-stakes issue in attosecond physics since the returning electron, by carrying back the energy it has absorbed from the laser, can act as the agent of many key processes in intense laser physics~\cite{Agos10}, including the ultrafast imaging of macromolecules~\cite{Lein07} and the design of new light sources through generation of ultrahigh harmonics~\cite{Popm12}.  When the laser is linearly polarized, the ``recollision'' (or ``rescattering'')~\cite{Cork93,Scha93} model has been immensely successful in both interpreting current experiments and devising new ones. Called the ``keystone of strong-field physics"~\cite{Beck08}, it describes how an ionized electron returns to the core in the next half laser cycle to share the energy it has acquired from the field with its parent ion. This energy may result in the emission of coherent radiation (high harmonic generation~\cite{Budi93,Band05}) or, if the energy exchange is sufficient, it may alter the core structure, e.g., leading to nonsequential double (or even multiple) ionization (NSDI) or fragmentation in molecules~\cite{Guo98, Guo01}. For linear polarization, an experimental signature of NSDI, and thus recollision, is the characteristic ``knee'' shape in the double ionization yield versus laser intensity. The arguments explaining recollision with linear polarization also predict its absence in a circularly polarized (CP) field because ionized electrons tend to spiral away from the core~\cite{Cork11}. When a knee was found for magnesium with a CP field~\cite{Gill01} a decade ago, recollision was immediately ruled out as a possible explanation. Recently, we reconciled this surprise with other experimental results in CP where no knee was observed~\cite{Fitt94}, by showing that recollision is possible with CP~\cite{Maug10_3}. 

In this Letter we show that recollisions result from a subtle compromise between the action of the strong laser field (which leads the electron away in a swirling motion) and the Coulomb attraction (which tends to recall it). Special families of orbits, which we call recolliding periodic orbits (R-PO), turn out to be this compromise:  They drive the recollision process by repeatedly ionizing and returning to the core. Considered together with an energy condition to alter the core structure, R-POs give rise to a few useful rules-of-thumb which can predict whether or not a target species (atom, molecule, or cluster) will exhibit NSDI, and if so, at which intensities. 

\begin{figure}
  \centering
  \includegraphics[width = 	\linewidth, type=png, read=.png, ext=.png]{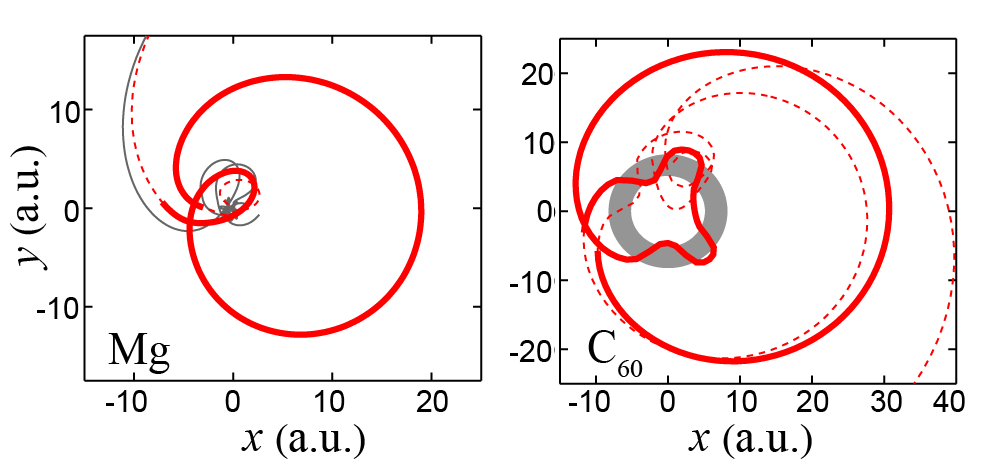}
  \caption{\label{fig:sampleReturns} 
  (color online) Left panel: Typical NSDI of Mg as modeled by Hamiltonian~(\ref{eq:Ham_2e2d}).  The dark (red online) electron exhibits a recollision which causes the gray electron to ionize.  Right panel: Typical one-electron trajectory of ${\rm C}_{60}$ which also exhibits a recollision~\cite{Hert05}. The gray annulus represents the potential well. In both panels, the segment of the trajectory which is in bold mimics the shape of key periodic orbits.  The laser parameters are 780~nm wavelength and an intensity of $5 \times 10^{13}\: \rm{W} \cdot \rm{cm}^{-2}$.  All trajectories are shown in the rotating frame.}
\end{figure}

\begin{figure}
  \centering
  \includegraphics[width = \linewidth, type=png, read=.png, ext=.png]{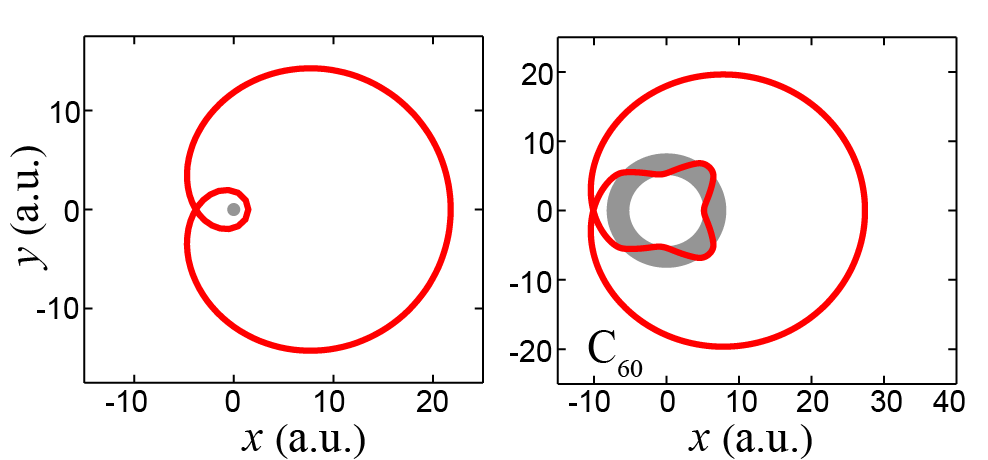}
  \caption{ \label{fig:RPO}
    (color online) R-PO corresponding to the highlighted trajectory portions in Fig.~\ref{fig:sampleReturns}. The left panel corresponds to the hard Coulomb potential, as given by Hamiltonian~(\ref{eq:Ham_rot}) and the right panel corresponds to the one-electron Hamiltonian model for $C_{60}$~\cite{Hert05}.}
\end{figure}

A typical NSDI trajectory in a CP field for a two-active-electron atom appears in the left panel of Fig.~\ref{fig:sampleReturns}: The dark electron (red online) recollides and causes the other electron (light gray), previously bound to the nucleus, to ionize. For comparison, the right panel of Fig.~\ref{fig:sampleReturns} displays a recolliding trajectory for ${\rm C}_{60}$ (buckminsterfullerene), where the gray annulus represents the potential well of the molecule. In both panels, we highlight the R-PO signature in bold.
We compare these recolliding trajectories with the periodic orbits of an already-ionized electron in Fig.~\ref{fig:RPO} for each system respectively: The good agreement between the shapes of the R-POs of Fig.~\ref{fig:RPO} and the sample trajectories of Fig.~\ref{fig:sampleReturns} (bold portions) forms the nub of our argument that specific periodic orbits drive recollisions.

The trajectories of Figs.~\ref{fig:sampleReturns} and \ref{fig:RPO} are represented in a frame rotating with the CP field (referred to as the rotating frame). They each consist of an interior loop, occurring in the down-field direction of the laser, which passes close to the nucleus (where the Coulomb attraction dominates the dynamics) and a farther reaching exterior loop which encloses the nucleus (where the laser field predominates). Comparing the R-PO in Fig.~\ref{fig:RPO} for Hamiltonian~(\ref{eq:Ham_rot}) (left panel) and  the same R-PO for ${\rm C}_{60}$ (right panel), we observe striking similarities, despite the strong differences in the potentials. The only common feature between the two potentials is the Coulomb tail far from the core ($-1/r$ for $r\gg1$) which results in a one electron model which we examine next.    
 
\begin{figure}
  \centering
  \includegraphics[width = \linewidth, type=png, read=.png, ext=.png]{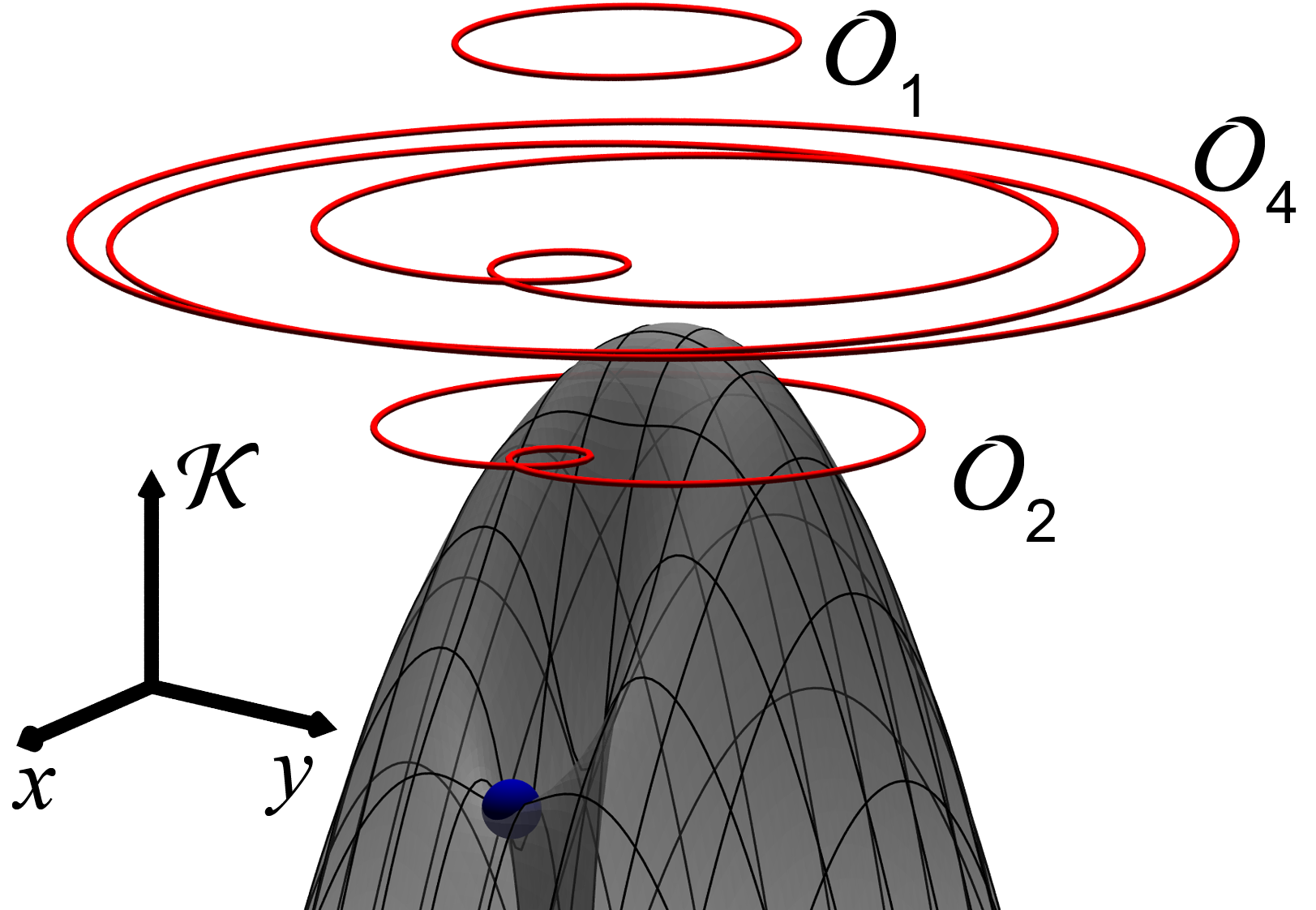}
  \caption{\label{fig:zvs} (color online) Limits of the accessible domain for an electron modeled with Hamiltonian~(\ref{eq:Ham_rot}). We display three R-POs of the family $\mathcal{O}_n$. The saddle point is marked by a sphere. The laser parameters are the same as in Fig.~\ref{fig:sampleReturns}.}
\end{figure}

In the rotating frame the Hamiltonian modeling a one-electron dynamics reads
\begin{equation}\label{eq:Ham_rot}
   \mathcal{K}=\frac{p_x^2+p_y^2}{2}-\frac{1}{\sqrt{x^2+y^2}} + E_0  x -\omega\left(xp_y - yp_x\right),
\end{equation}%
where the energy $\mathcal{K}$ is referred to as the Jacobi constant for its link to celestial mechanics~\cite{Hill78}. The variables $(x,y)$ and $(p_x,p_y)$ are the position and canonically conjugate momentum of the electron. $E_{0}$ is the amplitude of the field and $\omega$ its frequency. For ${\rm C}_{60}$ we use a continuous approximation of the potential given in Ref.~\cite{Hert05}. The phase space of Hamiltonian~(\ref{eq:Ham_rot}) is unbounded, but not all electrons can leave the core region and ionize. In particular if the Jacobi constant of the electron is smaller than the one corresponding to a specific Stark saddle point~\cite{Clar85, Howa95, Farr95}, the electron is stuck in the core region with no possibility for ionization. In Fig.~\ref{fig:zvs}, we display the limits of the domain accessible to the electron (in configuration space), where the saddle point is indicated by a sphere. 

As illustrated by Figs.~\ref{fig:sampleReturns} and~\ref{fig:RPO}, recollisions with CP are organized by certain types of periodic orbits. Only those periodic orbits which contain segments both close to and far away from the core play a role in recollisions and these are R-POs. The topologically simplest R-PO consists of an off-centered circle and we found such an orbit for high values of the Jacobi constant (top orbit in Fig.~\ref{fig:zvs}).  When followed by continuity as the Jacobi value is decreased, this orbit bifurcates into a family of periodic orbits which consist of an interior loop, in the down-field direction of the laser, and one or several exterior loops, leading to an excursion far from the core (like the highlighted regions of Fig.~\ref{fig:sampleReturns}). We label these R-POs $\mathcal{O}_{n}$, where $n$ corresponds to the number of loops in the periodic orbit (see Fig.~\ref{fig:zvs}). The interior loop is responsible for the exchange of energy between the already-ionized electron and the core (or second) electron while the exterior loop is where the ionized electron gains energy from the laser field.  Not all R-POs are equally important for recollision: Relevant orbits have a period which is much smaller than the pulse duration (so as to influence the motion of the already-ionized electron), and are weakly hyperbolic, so that an electron can stay close to them long enough to imitate their dynamics. We found a handful of such R-POs and their influence waxes and wanes with the choice of Jacobi value and intensity. For instance, the influence of $\mathcal{O}_{4}$ (middle curve in Fig.~\ref{fig:zvs}) can be seen on Fig.~2 of Ref.~\cite{Maug10_3}. 

\begin{figure}
  \centering
  \includegraphics[width = \linewidth, type=png, read=.png, ext=.png]{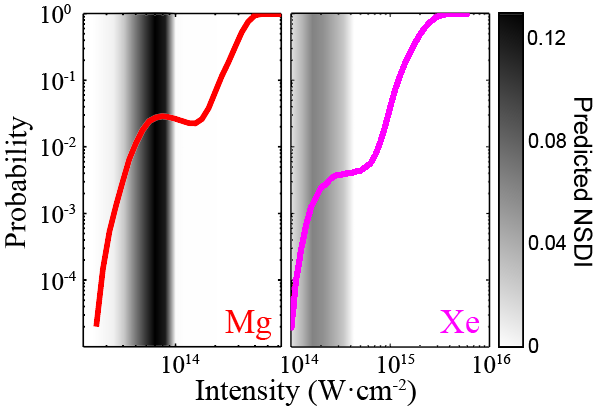}
 \caption{\label{fig:probDI}
   (color online) Curves: Double ionization yield for Mg (left panel) and Xe (right panel), obtained from Hamiltonian~(\ref{eq:Ham_2e2d}). In both panels the gray shaded background corresponds to the probability that the three criteria for NSDI are satisfied. The laser wavelength is $780$~nm.}
\end{figure}

The main outcome of the recollision is the modification of the core structure, leading to, e.g., NSDI. Recollisions exhibited by the two-electron Hamiltonian follow the organizing structures of the one-electron Hamiltonian (as seen in Fig.~\ref{fig:sampleReturns}) and indeed it will be shown that these structures dictate the properties of the NSDI channel. The two-electron dynamics can be expressed by~\cite{Beck12}
\begin{eqnarray}
    \mathcal{H} & = & \frac{\left|{\bf p}_1\right|^2+\left|{\bf p}_2\right|^2}{2}
                     +V\left(\left|\bf{r}_1\right|\right) + V\left(\left|\bf{r}_2\right|\right) + 
                     \frac{1}{\sqrt{\left|{\bf r}_1-{\bf r}_2\right|^2+b^2}} \nonumber \\
					     &   & + E_0 f(t) \left({\bf r}_1+{\bf r}_2\right)\cdot
					           \left(\begin{array}{c} \sin\omega t \\ \cos\omega t \end{array}\right). \label{eq:Ham_2e2d}
\end{eqnarray}%
Here ${\bf r}_{1,2}$ and ${\bf p}_{1,2}$ are the canonically conjugate positions and momenta of the two electrons in the lab frame. The potential is chosen as $V(r)=-2/\sqrt{r^2+a^2}$, where $a$ is the electron-core softening parameter~\cite{Java88, Panf01} which is adjusted to model the various atoms under investigation (for Mg, we set $a=3$, for Ne, $a=1$, for Ar, $a=1.5$, and for Xe, $a=1.8$). The pulse envelope is given by $f$. For the numerical results presented here, we consider a laser envelope with two laser cycle ramp-up and six cycle plateau. We find qualitatively similar results with other laser envelopes. Initial conditions of the respective atoms are taken from a microcanonical distribution on the ground state energy (defined as the sum of the two first ionization potentials. In this Letter we restrict Hamiltonian~(\ref{eq:Ham_2e2d}) to two spatial dimensions (fully three dimensional calculations follow the same organizational structure). In Fig.~\ref{fig:probDI} we show the probability of double ionization for Mg (left panel) and Xe (right panel), where we use a distance criterion for ionization.  Both atoms exhibit NSDI, which manifests itself in the knee enhancement. It should be noted that for Ar and Ne, such knees are absent in the double ionization probability curve versus intensity. In all cases, trajectory inspection shows that NSDI corresponds to recolliding trajectories.

\begin{figure}
  \centering
  \includegraphics[width = \linewidth, type=png, read=.png, ext=.png]{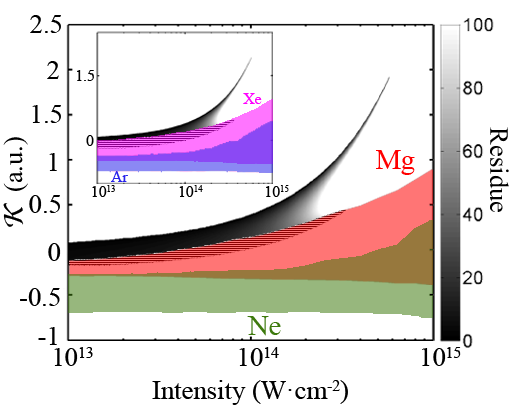}
  \caption{\label{fig:bifDiagram}
  (color online) Gray surface: Domain of existence of the R-PO $\mathcal{O}_2$ for Hamiltonian~(\ref{eq:Ham_rot}).  The gray shading corresponds to the Greene's residue~\cite{Gree79}, an indication of the stability of ${\mathcal O}_2$. Colored surfaces: Distributions of Jacobi values of the pre-ionized electron for Mg (red) and Ne (green).  Inset: Same as the main panel with Xe (magenta) and Ar (blue).  In both panels, the hatched areas correspond to the overlap of the Jacobi distributions with the domain of existence of ${\mathcal O}_2$.  The laser wavelength is 780~nm.}
\end{figure}

In order to interpret the presence of a knee enhancement in the double ionization probability versus laser intensity, we consider the recollision mechanism driven by R-POs. Given that R-POs are well defined for a constant laser envelope, we consider $f=1$ and address recollisions happening during the plateau for the two-electron calculations. It should be noted that the analysis extends to events happening during the ramp-up, by considering a lower effective laser intensity. In order to be influenced by a R-PO, a trajectory should get close to it and therefore have a Jacobi value compatible with those of the R-PO family. The analysis of the recollision dynamics shows that the existence of an overlap between the Jacobi values for the pre-ionized electron and the domain of existence of the R-POs accurately predicts the existence of recollisions. From this observation, we derive a simplified predictive model for the existence of recollision and NSDI for a given atom. Given the similarity between the R-POs with different models, we use Hamiltonian~(\ref{eq:Ham_rot}) to determine the domain of existence of R-POs and restrict the analysis to $\mathcal{O}_{2}$, irrespective of the atom. Due to the turn-on of the field the energy gained by each electron is $E_0x_i^{\left(0\right)}$ where $x_i^{\left(0\right)}$ is the $x$-coordinate of the $i^{th}$ electron at the beginning of the plateau. The generated distribution of Jacobi values corresponds to the colored areas in Fig.~\ref{fig:bifDiagram}. The gray strip shows the domain of existence of the R-PO $\mathcal{O}_2$ in the parameter space $\left(I, \mathcal{K}\right)$ for Hamiltonian~(\ref{eq:Ham_rot}). Following the previous discussion, we estimate the probability of recollision as the proportion of pre-ionized electrons with the Jacobi values compatible with the existence of the R-PO $\mathcal{O}_{2}$. Visually, it corresponds to the overlap between the gray and respective colored regions in Fig.~\ref{fig:bifDiagram} (hatched area).  There is significant overlap between the surfaces for Mg and Xe and the domain of existence of $\mathcal{O}_2$, therefore recollisions are expected for both atoms (at intensities where there is overlap). In contrast, Ne and Ar do not show any overlap so recollisions are not expected for these atoms in the near infrared regime. These predictions are confirmed by two-electron simulations of Hamiltonian~(\ref{eq:Ham_2e2d}) for the corresponding atomic models. Other R-POs, e.g., $\mathcal{O}_3$ and $\mathcal{O}_4$, can be included in the analysis. However, the results are very robust and do not change quantitatively because of the strong overlap in the domains of existence of the individual $\mathcal{O}_n$ in the parameter space $\left(I, \mathcal{K}\right)$.  %

Not every recollision leads to NSDI. For that to happen, the returning electron needs to bring sufficient energy to the core region in order to ionize the bound electron while remaining ionized itself. Based on this simple picture, we further refine the recollision criterion to predict the range of intensities where NSDI, and thus the knee in double ionization versus laser intensity, is expected. For a given recolliding trajectory, we define $\mathcal{K}_{1}$ (resp. $\mathcal{K}_{2}$) and $\mathcal{K}_{1}^{\prime}$ (resp. $\mathcal{K}_{2}^{\prime}$) as the Jacobi values of the pre-ionized (resp. core) electron before and after the recollision and we define $\eta^{2}$ as the energy exchange. Since the core electron is not screened by the pre-ionized electron, an effective charge of $-2$ is used to compute its Jacobi value from Hamiltonian~(\ref{eq:Ham_rot}). Assuming elastic recollisions leading to double ionization, the following inequalities hold:
\begin{subequations}\label{eq:energyExchange}
	\begin{eqnarray}
		\mathcal{K}_{1}^{\prime} = \mathcal{K}_{1}-\eta^2&>\mathcal{K}_\star, \\ 
	  	\mathcal{K}_{2}^{\prime} = \mathcal{K}_{2}+\eta^2&>\mathcal{K}_\star,
	\end{eqnarray}
\end{subequations}
where $\mathcal{K}_\star$ is the energy of the Stark saddle. Combining the two equations, we arrive at the condition $\mathcal{K}_{1}+\mathcal{K}_{2}>2 \mathcal{K}_\star$ which ensures that the returning electron is sufficiently energetic to trigger NSDI. In summary there are three necessary conditions for recollision-driven NSDI to occur:
\begin{enumerate}[(i)]
   \item \label{list1} the existence of an R-PO (in a single active electron approximation) for Hamiltonian~(\ref{eq:Ham_rot}),
   \item \label{list2} an overlap between the distribution of Jacobi values of the pre-ionized electron and the domain of existence of this R-PO, 
   \item \label{list3} the pre-ionized electron brings in a sufficient amount of energy to free the second electron.
\end{enumerate}
The first two conditions ensure that recollisions are possible, while the third criterion ensures that recollisions could lead to NSDI.  Varying the intensity and estimating the probability of these conditions provides the approximate intensity range where NSDI is possible.  Referring back to Fig.~\ref{fig:probDI} we compare these probabilities (gray surface) for  Mg (left panel) and Xe (right panel) with their respective double ionization curves, as given by Hamiltonian~(\ref{eq:Ham_2e2d}). We see that the intensity range at which NSDI occurs, i.e., the location of the knee, is predicted well by the three conditions. Numerically, Xe shows NSDI at intensities approximately one order of magnitude larger than in Mg~\cite{Wang10_2} (see also Fig.~\ref{fig:probDI}), in agreement with the probability to satisfy the three conditions. These predictions also agree with experimental findings~\cite{Guo98,Gill01}.  Finally, since the double ionization curves of Fig.~\ref{fig:probDI} are computed with a ramp-up, the electron experiences an effectively lower intensity during the ramp-up.  This results in double ionization curves which exhibit a cut-off intensity for the NSDI channel which is slightly higher than what is 	predicted from the three criteria.

Our discussion on recollision-driven events boils down to a few rules-of-thumb which apply to systems more complex than atoms. Molecular recollisions with CP are usually attributed to the spatial extent of the system, where the pre-ionized electron recollides at a different atomic center than the one it originates from~\cite{Guo01}. In contrast, we have argued above that the excursion of the electron is much larger than the size of the molecule (see Fig.~\ref{fig:sampleReturns}), and therefore the possibility of recollision in a CP field for molecules is not due to their spatial extent but mainly to how easily the first electron -- the energy carrier -- can be pre-ionized by the laser.  For example, Ref.~\cite{Guo01} reports a knee for NO and none for ${\rm N}_2$ using a near-infrared CP field.  The first ionization potentials of ${\rm N}_2$ and Ar are close, whereas NO resembles Mg. Since there are no recollisions for Ar at this wavelength (see inset of Fig.~\ref{fig:bifDiagram}), none should be expected for ${\rm N}_2$. On the other hand, recollisions are expected for NO since they are plentiful in Mg. Furthermore, since the second electron is more tightly bound in NO than it is in Mg, these recollisions need to bring back more energy for double ionization and hence experimentally NSDI is seen at higher intensity for NO than for Mg~\cite{Guo01,Gill01}.

A.K.\ acknowledges financial support from the Chateaubriand fellowship program of the Embassy of France in the United States, and F.M.\ from the Fulbright program.
The research leading to these results has received funding from the People Program (Marie Curie Actions) of the European Union's Seventh Framework Program FP7/2007-2013/ under REA grant agreement 294974. We acknowledge funding from the US National Science Foundation grant PHY0968866. 


\end{document}